\documentclass{article}
\usepackage{epsfig}
\usepackage{a4,amsmath,amsthm,amsfonts,amscd,eucal,latexsym,amssymb}

\def\cA{{\cal A}}
\def\cB{{\cal B}}
\def\cC{{\cal C}}

\def\cH{{\cal H}}

\def\cP{{\cal P}}

\def\cR{{\cal R}}
\def\cS{{\cal S}}
\def\cT{{\cal T}}

\def\cX{{\cal X}}

\def\bC{{\mathbb C}}

\def\bR{{\mathbb R}}

\def\bZ{{\mathbb Z}}

\def\b{\beta}
        
        \def\D{\Delta}

\def\m{\mu}
\def\n{\nu}

\newsymbol\bt 1202  

\renewcommand{\o}{\omega}
\renewcommand{\O}{\Omega}
\renewcommand{\l}{\lambda}
\renewcommand{\L}{\Lambda}

\newcommand{\eins}{{\rm 1\!\!l}}

\begin{document}
\noindent
\begin{center}
{ \Large \bf Vacuum Fluctuations, Geometric Modular Action\\[6pt]
and Relativistic Quantum Information Theory}
\\[30pt]
{\large \sc
Rainer Verch}
\end{center}
${}$\\
\begin{center}
                 Institut f\"ur Theoretische Physik,\\
                 Universit\"at Leipzig,\\
                 Postfach 10 09 20,\\
                 D-04009 Leipzig, Germany\\
                 e-mail: verch$@$itp.uni-leipzig.de
\end{center}
${}$\\[22pt]
{\small
{\bf Abstract.} A summary of some lines of ideas leading to model-independent
frameworks of relativistic quantum field theory is given. It is followed
by a discussion of the Reeh-Schlieder theorem and  geometric modular
action of Tomita-Takesaki modular objects associated with the quantum field
vacuum state and certain algebras of observables. The
distillability concept, which is significant in specifying useful entanglement
in quantum information theory, is discussed within the setting of general relativistic quantum field theory.
  }
\section{Introduction}
About 100 years ago, new insights into the physical world were gained which at
that time had a new quality to them. The new feature was that certain phenomena
could successfully be described by means of concepts which have little in common
with the behaviour of physical objects familiar from everyday expericence.
The first of these insights we are referring to was Planck's quantum hypothesis in his
account of black-body radiation. The second was Einstein's theory of special relativity.
(See, e.g., \cite{Pais} for a historical presentation of these developments.)
 \par
It took a while --- more or less, two decades --- until quantum theory reached the
form of (non-relativistic) quantum mechanics which is nowadays taught in courses
at universities. A further step was the combination and unification of the principles 
of quantum mechanics and special relativity. The endeavours to accomplish this step
took still longer --- and, rigorously speaking, they haven't come to an end even today.
And the synthesis of quantum mechanics and general relativity into some form of a
quantum theory of gravity lies still well ahead of us.
 \par
The theory unifying the principles of quantum mechanics and special relativity has
come to be called {\em relativistic quantum field theory}, or QFT, for short.
To delineate the basic characteristics of QFT, let us recall first the basic features
of
\\[10pt]
{\bf Quantum mechanics}, which provides a conceptual foundation for describing physical
processes at small scales (in space and time), and is therefore relevant in the 
microscopic domain and accounts for the stability of atoms and molecules. Moreover,
its (experimentally testable) predictions are of statistical nature, with the characteristic
feature of uncertainty relations.
\\[10pt]
{\bf Special relativity}, on the other hand, can be viewed as providing a conceptual foundation
for the description of space and time, relevant in particular in the context of
processes involving very high energies and momenta. Among its principal features are the
absence of preferred inertial frames (observers), i.e.\ Poincar\'e-covariance, the speed
of light as maximal velocity of signal propagation, and matter\,(mass)-energy equivalence.
\\[10pt]
The fundamental aspects of both quantum mechanics and special relativity find a unification
in the form of 
\\[10pt]
{\bf Quantum field theory}, which consequently provides a theoretical framework
for the description of processes with very high energy/momentum exchange at very
small time/length scales; it is therefore relevant in the sub-microscopic domain
and accounts for the properties and the stability of elementary particles,
predicts annihilation and creation of particles, new types of charges, anti-charges,
PCT and spin-statistics theorems, fluctuations and long-range correlations.
\\[10pt]
While this is not the place to give a review of the historical development of QFT
and its interplay with the development of elementary particle physics,
involving also new concepts such as renormalization, internal group symmetries, gauge theory,
spontaneous symmetry breaking, Higgs mechanism etc.,
there are some comments to be made at this point about the various sub-branches
of QFT and its status as a physical theory, as well as its status as concerns mathematical
consistency of the framework. 
 \par
Let us begin by mentioning the by far largest branch
of QFT, which we refer to as {\em perturbative QFT}. The idea here is to look at concrete
{\em quantum field models}, mostly in the form of a Lagrangean for an --- initially ---
classical field theory model involving certain types of matter and gauge fields.
Typically, the fields interact in some way and this leads to the occurrence of
multilinear (polynomial) expressions of the fields in the field equations.
One would then like to have ``quantized'' solutions of the field equations. It is
not a priori clear what this means, but the pragmatic way to proceed is as follows. 
One starts with the interaction-free part of the field equation (neglecting the
multilinear, interacting parts of the field equations) and constructs ``quantized''
solutions for that in the form of ``free'' quantum fields --- where it is in most of the
relevant cases known what this means. Then one regards the interacting expressions
of the (now quantized, free) fields as a perturbation of the free dynamics, and tries
to construct solutions to the full dynamics by means of a perturbation series in the
parameter specifying the strength of the interaction (the coupling parameter). At this
point there arises the difficulty that the various multilinear expressions in the
fields appearing in the perturbation series
are not well defined at the level of (free) quantized fields, and that they
need to be ``renormalized''. If this is possible systematically to all polynomial
orders upon introducing only finitely many parameters (to be determined experimentally),
one calls the quantum field model under consideration {\em (perturbatively) renormalizable}.
Once the renormalization parameters are determined experimentally, predictions of the
quantum field model can be compared with experimental data e.g.\ obtained in scattering
experiments with elementary particles --- up to a given order
in the coupling parameter of the perturbation series.
 \par
The successes of perturbative quantum field theory in comparison with experiment are
truely impressive. The numerical agreement of theoretical predictions and experimental data is
in many cases
of the in the range of 8 significant figures or better, and also properties of 
particles whose existence was
predicted by QFT prior to observation, like in the case of the ${\rm W}^\pm$ and 
${\rm Z}^0$ bosons in the
electroweak interactions, are in excellent agreement with experimental findings.
 (See \cite{Weinberg,ItzZub} for the various aspects of perturbative
QFT.) 
 \par
However, from a more fundamental point of view, perturbative QFT is not fully satisfactory.
The perturbation series by which one attempts to approximate the full interacting quantum
field dynamics won't converge, and then it is unclear if there is a solution to the
quantized field equations at all. This provokes the question at which order 
in the coupling parameter the perturbation
series ought to be truncated to yield acceptable agreement with experiment, and this 
question remains
so far unanswered within perturbative quantum field theory.    
Moreover, the number of renormalization parameters which have to be determined by experiment
and are not derivable within perturbation theory are quite large for physically realistic
quantum field models (of the order of about 20 in the case of the standard model), and this
is regarded as a considerable drawback as concerns the predictive power of perturbative QFT.
 \par
Hence, there clearly is room for approaches to QFT (and elementary particle physics) other
than by perturbative QFT. Let me point out three basic branches. One idea is that theories
such as the standard model are simply not rich enough and/or do not include all interactions
(such as gravity), and that a richer theory should be considered in the first place (first
at the level of a ``classical field theory''
 then quantized, maybe at the level of perturbative QFT), with the hope that
the richer symmetry structure constrains the amount of free parameters considerably.
{\it Grand unified theories}, and {\it string theory}, can be seen in this light.
 \par
The next branch is {\it constructive quantum field theory}, where one attempts to construct
solutions to the quantized, interacting field equations mathematically rigorously. This
branch of QFT is much smaller than those mentioned previously, but has had quite impressive
successes which are partly documented in \cite{GlimmJaffe,Rivasseau}. The mathematical
difficulties one is faced with in constructive QFT are immense, not least by the circumstance
that it is often not entirely clear what is actually meant by a solution to a quantized, 
interacting field equation (we will soon come back to this point).
 Nevertheless, interacting quantum field models have been rigorously
constructed in spacetime dimensions 2 and 3. The case of a rigorous solution to quantized
field equations for models regarded as physically relevant remains open in 4
spacetime dimensions
and is still an area of active reasearch. The Clay Institute of Mathematics awards a million
dollars for the solution of this problem. [There is also a branch of QFT which
is known as {\em lattice gauge theory}, and which can be placed somewhere between perturbative
QFT and constructive QFT. The interested reader is referred to \cite{MontvayMunster} for
more information about it.]
 \par
Finally, there is yet another branch of QFT, commonly called 
{\em axiomatic quantum field theory},
although this labelling is to some degree misleading. The basic idea is that one wishes to
formulate and 
analyze the properties  which are thought to be common to all physically realistic quantum
field models. This is on one hand indispensible to make the problem of rigorous construction
of interacting quantum field models a mathematically well-defined problem, on the other hand
it is also difficult in the absence of rigorously constructed interacting 
quantum field models in 4 spacetime
dimensions as a guidance. To begin with,
 the task is to find a mathematical structure which encodes the
basic principles of quantum mechanics and special relativity, and which subsumes the known
rigorously constructed quantum field models where these principles are implemented (e.g.\
for free quantum fields, or for interacting quantum fields in lower spacetime dimension).
This task was taken up initially by Wightman and others (see \cite{StrWigh,Jost,BLOT})
 from a point of view
involving mainly distribution theory, and by Haag and Kastler \cite{HK,Haag} using
the mathematical theory of operator algebras. Seen from a mathematician's perspective,
the latter approach turned out to be more fruitful. In fact, there are many rigorous
 and deep results
about the mathematical structure of (model-independent) quantum field theory in the operator
algebraic framework. The reader might like to consult \cite{Haag,Araki,Baumg} for
 a comprehensive review. 
 \par
The present contribution is, in fact, placed within the framework of axiomatic QFT.  
In the next section, we will sketch how one can combine the principles of quantum
mechanics and of special relativity in a mathematical structure which more or less
is ``common to all quantum field models''. Then we will present the ``Reeh-Schlieder-theorem''
and discuss some of aspects of it. The Reeh-Schlieder-theorem is a strong mathematical statement
about the ubiquity and complexity of vacuum fluctuations in quantum field theory, regardless
of the particular quantum field theoretical model considered: It is a consequence of first
principles such as locality (causal propagation), stability of the vacuum, and covariance.
Then we will discuss a mathematical structure arising in connection with the Reeh-Schlieder-theorem:
Geometric modular action. While discovered already in 1975 by 
Bisognano and Wichmann \cite{BiWi}, this
mathematical structure has in the recent years given rise to many new insights into
quantum field field theory which we will briefly discuss. In a sense, it unifies the 
mathematical domains of quantum mechanics --- operator algebras --- and of special relativity
--- affine geometry --- completely. Moreover, it opens very interesting new perspectives.
 \par 
We will then proceed to another topic where the Reeh-Schlieder-theorem plays again a
prominent role: In discussing aspects of entanglement in the framework of relativistic
QFT. This part of the present contribution is essentially a summary of parts of a recent joint
work with R.\ Werner \cite{VerWer}. We will present a variant of the distillability concept
of bipartite systems in quantum field theory. Furthermore, we will
quote our result stating that the vacuum state
(as well as any relativistic thermal equilibrium state) 
is distillable over arbitrary spacelike distances.
 \par
Taking up a line of thought mentioned at the very beginning of this introduction, we should
like to point out that also in the realm of phenomena described by quantum field theory 
one encounters theoretical propositions which at first sight appear implausible because of
their highly counterintuitive character. The Reeh-Schlieder-theorem serves as an example,
as well as distillability of the vacuum state. However, careful statement of the concepts
and careful analysis of their consequences, together with proper use of adequate
mathematical methods, will bring us closer to an understanding of these novel situations
and, ultimately, their experimental testing. Thus, we will need to collect also some
mathematical concepts and results which are not necessarily in every theoretical physicist's
toolbox. Nevertheless, we have tried to keep the amount of formalities at a minimum and
to make this contribution as self-contained as possible, hoping that everyone familiar with
quantum mechanics, special relativity and the rudiments of quantum field theory will be 
able to follow this contribution without undue strain.
\section{From Quantum Mechanics and Special Relativity to Quantum Field Theory}
Let us once more recall the basic features of quantum mechanics, this time at a more formal
level. The theory of quantum mechanics says that a quantum mechanical system is
described by:
\\[10pt]
$\Box$ \quad $\cH$ : a Hilbert space\\[10pt]
$\Box$ \quad $\cR \subset B(\cH)$ : a $*$-algebra of operators, where:
\\[10pt]
${}$ \quad \ \ $\bullet$ $A = A^* \in \cR$ is interpreted as an {\it observable}
\\[10pt]
${}$ \quad \ \ $\bullet$ For $\psi \in \cH$ with $||\psi|| = 1$, the quantity
$$ \langle A \rangle_\psi = \langle \psi,A\psi \rangle $$ 
is interpreted as the {\it  expectation value} of the observable $A$ in the {\it  state}
given by $\psi$.
More generally: For $\rho = \text{trace-class operator on}\ \cH$ with
$\rho \ge 0$, $\text{\rm trace}(\rho) = 1$, we interpret $\langle A \rangle_\rho = \text{\rm trace}(\rho\,A)$
as expectation value of $A$ in the state given by $\rho$.
\\[10pt]
We need to explain some notation and terminology appearing here.
First note that by Hilbert space we mean a complex-linear Hilbert space. 
The scalar product of two vectors $\psi,\phi \in \cH$ is denoted
$\langle \psi,\phi \rangle$, and $||\psi||^2 = \langle \psi,\psi \rangle$. By $B(\cH)$
we denote the set of all bounded linear operators $A: \cH \to \cH$. A subset $\cR$ of
$B(\cH)$ (which may, but need not, coincide with $B(\cH)$) is a $*$-algebra if, given
$A$ and $B$ in $\cR$ and $\l,\m \in \bC$, the operators $\l A + \m B$, $AB$ and
$A^*$  are again cointained in $\cR$, where $A^*$ is the adjoint operator. Hence, a
quantum mechanical system is described by specifying its state Hilbert space $\cH$ and
its algebra of observables $\cR$.
 \par 
There are a few remarks to be made:
\\[6pt]
(2.1) One might take the point of view that the description of a quantum mechanical
system requires also the specification of dynamics, e.g.\ in the form of a Hamiltonian
operator $H$ acting in $\cH$. Furthermore, one may also require that the quantum system admits
states of lowest energy for $H$ (``ground states'') [or that the spectrum of $H$ is bounded
below], or thermal equilibrium states, since the sudden decay of matter which would otherwise
occur (for quantum systems not having these properties) is not observed in real systems.
We shall ignore aspects of dynamics for the moment, but will come back to this point
later in the discussion of quantum field theory.
\\[6pt]
(2.2) It is tacitly assumed that $\cR$ is non-abelian, i.e.\ that $AB \ne BA$ 
holds for some $A$ and $B$ in $\cR$, as otherwise there are no uncertainty
relations which are characteristic of quantum theory.
\\[6pt]
(2.3) One may wonder if the setting presented here is general enough since $\cR$
contains only bounde operators, while in quantum meachnics of single particles 
observables like position or momentum are represented by unbounded operators as a 
consequence of the canonical commutation relations. Employing the functional calculus,
however, one may pass e.g.\ from the unbounded operator $P$ representing the observable 
``momentum'' to the bounded operator $f(P)$, which is bounded when $f$ is a bounded
real function on $\cR$, and which represents the observable ``$f$(momentum)''. This 
shows that it is in general no loss of physical information to work only with 
bounde operators as observables; moreover, unbounded operators can be regarded as suitable
limits of sequences of bounded operators. Working with bounded operators has
considerable advantages as far as the mathematical analysis is concerned, since
subtle domain problems that plague the rigorous manipulation of unbounded operators
are avoided.
\\[6pt]
(2.4) One may also wonder why we have not simply taken $\cR = B(\cH)$, the standard case
in quantum mechnics of a single particle. The reason is that we would like to allow
greater flexibility, making it possible to consider also subsystems of a larger,
ambient system. An example, occuring often in quantum information theory, is the case
$\cH = \cH_1 \otimes \cH_2$ with $\cR = B(\cH_1) \otimes {\bf 1}$ modelling a subsystem
of the full system whose algebra of observables is given by $B(\cH_1 \otimes \cH_2) 
\backsimeq
B(\cH_1) \otimes B(\cH_2)$. We will encounter a similar situation later. In discussions
of model-independent properties of quantum field theories, $\cR$ often means the algebra
of observables measurable --- and in this sense, localized --- in a proper subregion
of Minkowski spacetime, as we will discuss below.
\\[10pt]
Having thus collected the basics of the formal framework of quantum mechanics, we turn now
to special relativity. We will be very brief in recalling its basic formal ingredients.
The theory of special relativity states that all physical events can be collected in a
catalogue which has the structure of a 4-dimensional affine space $M$, where each point
in $M$ represents a (possible) event.
There is a metric $\eta$ of Lorentzian signature on $M$; that is, one can choose identifications
of $M$ with $\bR^4$ in such a way that, with respect to the standard coordinates of
$\bR^4$, $\eta$ is represented by the diagonal matrix ${\rm diag}(1,-1,-1,-1)$. The choice
of such an identification is also referred to as fixing of an inertial frame. With respect to
a fixing of an inertial frame (inducing an orientation an a time-orientation), one can introduce
the proper orthochronous Poincar\'e group $\mathfrak{P}^\uparrow_+$, which is the unit connected
component of the full Poincar\'e group $\mathfrak{P}$, defined as the group of all invertible
affine transformations of $M$ leaving $\eta$ invariant.
We assume from now on that an inertial frame has been fixed. Any $L \in \mathfrak{P}$ (or
$\mathfrak{P}^\uparrow_+$) decomposes as a semidirect product of $\L \in \mathfrak{L}$
(or $\mathfrak{L}^\uparrow_+$), the Lorentz group (or its unit connected component) and
$a \in \mathfrak{T} \equiv \bR^4$, the group of translations, according to
$$ Lx = (\L,a)x = \L x + a\,, \quad x \in M \equiv \bR^4\,.$$
The reader is referred to the contribution by Domenico Giulini in this volume for a full
discussion of special relativity, Minkowski spacetime and the Poincar\'e group (alternatively,
see e.g.\ \cite{SexlUrbantke}). 
 \par
The theory of special relativity states that the description of a physical system is equivalent
for all inertial observers, i.e.\ in arbitrary inertial frames. Put differently, the description
of physical processes should be covariant with respect to proper, 
orthochronous Poincar\'e transformations.
More formally, this means:
\\[10pt]
Suppose a quantum system is modelled by $(\cR,\cH)$.
Let $\rho$ be a density matrix and $A$ an observable
with respect to a given inertial frame. If $L \in \mathfrak{P}^\uparrow_+$,
then there corresponds, with respect to the $L$-transformed inertial frame,
 a density matrix $\rho_L$ and observable $A_L$ to $\rho$ and $A$, respectively,
such that 
\setcounter{equation}{4}
\begin{equation} \label{covari}
 \langle A_L \rangle_{\rho_L} = \langle A \rangle_\rho\,.
\end{equation}
One can add some mathematical precision, requiring that the maps taking $A$ to $A_L$  and
$\rho$ to $\rho_L$ are one-to-one and onto, i.e.\ bijective. Following Wigner, one may think
of elementary systems where $\cR = B(\cH)$, and then one can conclude:
\\[10pt]
There is a unitary representation
$$ \widetilde{\mathfrak{P}}{}^\uparrow_+ \owns \widetilde{L} \mapsto \widetilde{U}(\widetilde{L}) $$
of the universal covering group of $\mathfrak{P}^\uparrow_+$ on $\cH$, such that
$$A_L = \widetilde{U}(\widetilde{L})A\widetilde{U}(\widetilde{L})^*\,, \quad \rho_L = \widetilde{U}(\widetilde{L})\rho\widetilde{U}(\widetilde{L})^*\,, $$
where $\widetilde{\mathfrak{P}}{}^\uparrow_+ 
 \owns \widetilde{L} \mapsto L \in \mathfrak{P}^\uparrow_+$ is the canonical projection.
Moreover, if suitable assumptions about the continuity of the maps $A \mapsto A_L$, 
$\rho \mapsto \rho_L$ are made --- and we tacitly make this assumption ---
 then one can conclude that the unitaries  $\widetilde{U}(\widetilde{L})$
depend continuously on $\widetilde{L}$.
\\[10pt]
This result is known as the Wigner-Bargmann-theorem, which actually holds under somewhat
weaker assumptions than expressed in \eqref{covari}; it is sufficient to consider as
observables 1-dimensional projections $A = |\psi \rangle \langle \psi|$ and likewise,
1-dimensional projections $\rho = |\phi \rangle \langle \phi|$ as density matrices, and to
replace \eqref{covari} by the weaker requirement
$$ |\langle A_L \rangle_{\rho_L}| = |\langle A \rangle_{\rho}|\,.$$
We refer to the original articles by Wigner \cite{Wigner} and Bargmann \cite{Bargmann} and
to \cite{Haag,StrWigh,RobRoep} for considerable further discussion. 
 \par
The Wigner-Bargmann-theorem states that, in the case of an (elementary) quantum
system compatible with the covariance principle of special relativity, the
state Hilbert space $\cH$ carries a unitary representation
of  $\widetilde{\mathfrak{P}}^\uparrow_+$, the universal covering group of the proper
orthochronous Poincar\'e group, implementing the change of inertial frames. The appearance
of  a unitary representation of the universal covering group instead
of the proper orthochronous Poincar\'e group itself is due to the fact that \eqref{covari}
fixes only a unitary representation of $\mathfrak{P}^\uparrow_+$ up to a phase, but this
can be lifted to a proper unitary representation of $\widetilde{\mathfrak{P}}^\uparrow_+$.
The significance of this was clarified by Wigner's analysis of the irreducible unitary
representations of $\widetilde{\mathfrak{P}}^\uparrow_+$ having positive energy, thereby making
the term ``elementary system'' precise. The Hilbert spaces supporting these irreducible
unitary representations (``one-particle spaces'') correspond to spaces of solutions of
linear wave equations, like the Klein-Gordon, Dirac or Maxwell equations in the simplest
cases. The mass and the spin (or helicity) of these wave equations is a distinguishing
label for the irreducible unitary representations of $\widetilde{\mathfrak{P}}^\uparrow_+$.
 \par
Wigner's analysis reveals some structural elements of quantum mechanical
systems compatible with the principles of special relativity, but not all,
in particular the aspect of a ``quantized field'' hasn't appeared yet.
To see how this aspect comes into play, one usually takes a complementary route:
Consider a typical $\mathfrak{P}^\uparrow_+$-covariant classical system; i.e.\ a classical
field subject to a linear wave-equation. The electromagnetic field provides
the prime and archetypical example, but let us consider here a much simpler
example, the scalar Klein-Gordon field $\varphi(x)$, $x \in M \equiv \bR^4$,
obeying the following equation of motion:
$$
    (\eta^{\mu\nu} \frac{\partial}{\partial x^\mu} \frac{\partial}{\partial x^\n} + m^2) 
 \varphi(x) = 0 $$
where $m \ge 0$ is a constant. Such a classical field can be viewed as a Hamiltonian
system with infinitely many degrees of freedom, and one may therefore try and quantize
it by regarding it as a ``limit'' of a Hamiltonian system with $N$ degrees of freedom
as $N \to \infty$, and taking as its quantized version the
``limit'' of the quantized systems with $N$ degrees of freedom as $N \to \infty$. In the
case of the Klein-Gordon field, the classical field $\varphi(x^0,\boldsymbol{x})$,
 $x = (x^0,\boldsymbol{x}) \in \bR \times \bR^3$, at time-coordinate $x^0$ (with
respect to an arbitrary but fixed inertial frame) can be approximated by a discrete lattice
of coupled harmonic oszillators with canonical coordinates 
$q_{\l\mu\nu}(x^0)$ at the lattice site
 $$\boldsymbol{x}(\l,\mu,\nu) =
a \left[\begin{array}{c} \l \\ \mu \\ \nu \end{array} \right] \in \bR^3\,, \quad 
\l,\mu,\nu \in \bZ\,, \quad |\l |, |\mu |, |\nu | \le \frac{1}{a^2}\,,$$
where $a > 0$ is the lattice spacing. To the discrete lattice system one can
associate the quantum system of coupled harmonic oszillators (at lattice spacing
$a$, there are $N \sim 1/a^6$ of them), where the canonical classical coordinates
$q_{\l\mu\nu}(x^0)$ and conjugate momenta $p_{\l\mu\nu}(x^0)$ become operators
$Q_{\l\mu\nu}(x^0)$ and $P_{\l\mu\nu}(x^0)$ obeying the canonical commutation
relations.
In the limit as $a \to 0$ and $N \to \infty$, one obtains for each $f,h \in
C_0^\infty(\bR^3)$ the field operators
\begin{eqnarray*}
                    \Phi(x^0,f) & = & \lim_{a\to 0,\, N \to \infty} \, \sum_{\l,\mu,\nu}
                               Q_{\l \mu\nu}(x^0)f(\boldsymbol{x}(\l,\mu,\nu)) 
                               a^3\,, \\
 \Pi(x^0,f) & = & \lim_{a\to 0,\, N \to \infty} \, \sum_{\l,\mu,\nu}
                               P_{\l\mu\nu}(x^0)f(\boldsymbol{x}(\l,\mu,\nu)) 
                               \,,\\
\end{eqnarray*}
 For a detailed discussion of this construction,
cf.\ \cite{HenleyThirring}. To summarize, we find the following formal 
correspondences (where we use the shorthand $j$ or $\ell$  for the index triple $\l\mu\nu$, and
occasionally drop the time-argument $x^0$):
\begin{center}
${}$\begin{tabular}{l|l}
{\bf Classical Mechanics} & {\bf  Quantum Mechanics}\\[6pt]
Phase-space fncts  & Operators \\
${}$\quad $q_1,\ldots,q_k,p_1,\ldots,q_k$ & ${}$\quad $Q_1,\ldots,Q_k,P_1,\ldots,P_k$
\\[6pt]
Poisson brackets: & Commutators: \\
${}$\quad $\{q_j,p_\ell\} = \delta_{j\ell}$ & ${}$\quad $[Q_j,P_\ell] = i\hbar \delta_{j\ell}$
\\[10pt]
{\bf Classical Field Theory} & {\bf  QFT} \\[6pt]
field \quad \quad can.\ conj.\ momentum &  \\
$\varphi(x^0,\boldsymbol{x})$ \quad \quad $\pi(x^0,\boldsymbol{x})$ & \\[6pt]
$\varphi(x^0,f ) = \int d^3\boldsymbol{x}\,f(\boldsymbol{x})\varphi(x^0,\boldsymbol{x})$
$f \in C_0^\infty(\mathbb{R}^3)$ & $\Phi(x^0,f)$\,,\ $\Pi(x^0,h)$:\ operators in $\cH$\\[6pt]
Approximation & \\
$\sum_j q_j(x^0)f(\boldsymbol{x}(j)) a^3 \to \int d^3\boldsymbol{x}\,\varphi(x^0,\boldsymbol{x})$ & \\[4pt]
$\Rightarrow$ Poisson brackets: & Commutators: \\
$\{\varphi(x^0,f),\pi(x^0,h)\} = \int  d^3\boldsymbol{x}\,f(\boldsymbol{x})h(\boldsymbol{x})$ &
 $[\Phi(x^0,f),\Pi(x^0,h)]
   = i \hbar \int  d^3\boldsymbol{x}\,f(\boldsymbol{x})h(\boldsymbol{x})$ 
\end{tabular}
\end{center}
So far we have introduced field operators $\Phi(x^0,f)$
and their canonically conjugate momenta $\Pi(x^0,h)$, at fixed 
inertial frame-coordinate time $x^0$. They are ``smeared'' against the 
spatial argument $\boldsymbol{x}$ with test-functions $f$ and $h$
in $C_0(\bR^3)$. Without smearing, the density-like quantities
$\Phi(x^0,\boldsymbol{x})$ and $\Pi(x^0,\boldsymbol{x})$ cannot be 
interpreted as operators on a Hilbert space as a consequence of the 
canonical commutation relations --- the entry in the lower right
corner of the just tabled scheme --- but only as quadratic forms. This is due to the distributional character of the $\Phi(x^0,\boldsymbol{x})$ and $\Pi(x^0,\boldsymbol{x})$, whence the
equal-time canonical commutation relations are often written in the form
$$ [\Phi(x^0,\boldsymbol{x}),\Pi(x^0,\boldsymbol{x}')] = i \hbar \delta(\boldsymbol{x}
 - \boldsymbol{x}')\,.$$
It is quite useful to introduce, for test-functions $F \in C_0^\infty(\bR^4)$ 
distributed over open subsets of Minkowski spacetime, the field operators
$$ \Phi(F) = \int d^4x \,F(x^0,\boldsymbol{x})\Phi(x^0,\boldsymbol{x}) =
 \int dx^0\,\Phi(x^0,f_{x^0})\,, \quad f_{x^0}(\boldsymbol{x}) = F(x^0,\boldsymbol{x})\,.$$
These field operators can be rigorously interpreted as unbounded (and for real-valued $F$,
selfadjoint) operators on a suitable domain of a Hilbert space $\cH$ which arises
as the bosonic Fock space over the one-particle space of solutions to the
Klein-Gordon equation with positive energy. This one-particle space carries an
irreducible, unitary representation of $\widetilde{\mathfrak{P}}{}^\uparrow_+$, which
lifts to a unitary representation of $\widetilde{\mathfrak{P}}{}^\uparrow_+$ on $\cH$.
Let us denote this representation by $U$, since it is actually a representation of
${\mathfrak{P}}{}^\uparrow_+$ in this case, as for every linear field equation of integer spin.
Then one finds that covariance holds in the form of
$$ U(L) \Phi(F) U(L)^* = \Phi(F \circ L^{-1})\,, \quad L \in {\mathfrak{P}}{}^\uparrow_+,\ \
F \in C_0^\infty(\bR^4)\,;$$
moreover, one also has 
\begin{equation} \label{qKG}
 \Phi((\eta^{\mu\nu} \frac{\partial}{\partial x^\mu} \frac{\partial}{\partial x^\nu}
+ m^2) F) = 0\,, \quad F \in C_0^\infty(\bR^4)\,,
\end{equation}
and there holds also the covariant form of the canonical commutation relations,
\begin{equation} \label{covCCR}
 [\Phi(F_1),\Phi(F_2)] = i\hbar G(F_1,F_2)\,, \quad F_1, F_2 \in C_0^\infty(\bR^4)\,,
\end{equation}
with the ``causal Green's function''
\begin{eqnarray*}
G(F_1,F_2) & = & {\rm Im} \,\int_{\bR^3} \frac{d^3\boldsymbol{p}}{\o(\boldsymbol{p})}
   \overline{\tilde{F_1}(\o(\boldsymbol{p}),-\boldsymbol{p})}\tilde{F_2}(\o(\boldsymbol{p}),-\boldsymbol{p}) \\
\o(\boldsymbol{p}) = \sqrt{\boldsymbol{p}^2 + m^2} \,, & & \tilde{F} = \text{Fourier-transform of}\ \  F
\end{eqnarray*}
which vanishes whenever the supports of $F_1$ and $F_2$ are causally separated.
 \par 
We shall not elaborate on the mathematical details related to the Fock space operators
$\Phi(F)$ since this is all well-documented in the literature
(see, e.g., \cite{ReedSimon2,BLOT}. Rather we should make the remark at this
point that the properties of the operators $\Phi(F)$, interpreted as Fock space
operators, may serve as a blue-print of a general concept of a (in this case,
scalar) ``quantum field'', as soon as they are abstracted from properties
pertaining to the model of the Klein-Gordon field, i.e., the equation of
motion \eqref{qKG}. The ensueing conceptual framework for a general scalar
quantum field are represented by the ``Wightman axioms'', which we list now,
not paying too much attention to full mathematical rigor (see \cite{StrWigh,BLOT,Jost}
for a more detailed exposition of these matters).
\begin{itemize}
\item[{\bf i)}] $\exists$ a Hilbert space $\cH$ with a dense domain $\mathcal{D} \subset \cH$, so that
     all $\Phi(F)$ are well-defined operators on $\mathcal{D}$, and 
     $\Phi(F)^* = \Phi(\overline{F})$ 
\item[{\bf ii)}] $F \mapsto \Phi(F)$ is complex linear and suitably continuous
\item[{\bf iii)}] {\bf  Covariance:} There is on $\cH$ a unitary representation\\[2pt]
 $\mathfrak{P}^\uparrow_+ \owns L \mapsto U(L)$, with $U(L)\mathcal{D} \subset \mathcal{D}$, so that
 $$ U(L) \Phi(F) U(L)^* = \Phi(F \circ L^{-1}) \quad \ \ (\ \Phi(x)_L = \Phi(L(x)) \ )$$
\item[{\bf iv)}] {\bf  Locality, or relativistic causality:} \\
If the {\it supports} of the test-function $F_1$ and $F_2$ are
{\it causally separated}, the corresponding field operators commute:
$$ [\Phi(F_1),\Phi(F_2)] = 0 $$
\item[{\bf v)}] {\bf  Spectrum condition/positivity of the total energy:}\\
Writing $U(1,a) = e^{iP_\mu a^\mu}$, it holds (in the sense of expectation values) that
$$ P_0^2 - P_1^2 - P_2^2 - P_3^2 \ge 0 \,, \quad P_0 \ge 0$$
\item[{\bf vi)}] {\bf  Existence (and uniqueness) of the vacuum:}\\
 $\exists$ $\Omega \in \mathcal{D}$, $||\Omega|| = 1$, so that $U(L)\Omega = \Omega$ and this
vector is uniquely determined up to a phase factor.
\item[{\bf vii)}] {\bf  Cyclicity of the vacuum:} \\
The domain $\mathcal{D}$ is spanned by vectors of the form
$$ \Omega,\ \Phi(F)\Omega,\ \Phi(F_1)\Phi(F_2)\Omega,\ldots, \ \Phi(F_1)\cdots\Phi(F_n)\Omega,\ldots $$
\end{itemize}
As indicated above, the just given collection of conditions tries to
capture the essential properties of a ``quantum field''. We notice that,
compared to the properties of the Klein-Gordon field, the commutation relations
\eqref{covCCR} have been generalized to the condition of spaccelike commutativity,
and the reference to a specific field equation has been dropped.
Spacelike commutativity says that there should be no uncertainty relations
between observables measured at causal separation from each other, and thus
gives expression to the principle that there is no operational signal propagation
faster than the speed of light. It should be remarked here that there is no
difficulty in generalizing the above stated conditions to fields of general
spinor- or tensor-type \cite{StrWigh,BLOT,Jost}. The basic difference is that
for fields of half-integer spin, spacelike commutativity of the field operators
must be replaced by spacelike anti-commutativity in order to ensure consistency with the
other conditions: This is, basically, the content of the spin-statistics theorem.
In this sense, a field carrying half-integer spin does not have the character of
an observable --- typically, it also transforms non-trivially under gauge
transformations. Observable quantities, and related quantum field operators
fulfilling spacelike commutation relations, can be built from half-integer spin
quantum fields by forming suitable bilinear expressions in those fields.
Once more, we must refer to the literature for a fuller discussion of these matters
\cite{BLOT}. 
 \par
Furthermore, it is worth noting that the type of Poincar\'e covariance {\bf iii}),
implemented by a unitary representation of ${\mathfrak{P}}{}^\uparrow_+$,
makes an explicit appearance here, completely in the spirit of the
Wigner-Bargmann theorem. (For fields of half-integer spin type, this must
be replaced by a unitary representation of  $\widetilde{\mathfrak{P}}{}^\uparrow_+$,
in keeping with the circumstance that such fields are not directly
observable.)
 \par
Some new aspect appears here which we have already alluded to in remark (2.1)
and which made an implicit appearance elsewhere when we referred to
irreducible unitary representations of $\widetilde{\mathfrak{P}}{}^\uparrow_+$
having {\it positive energy}. This is the aspect that the time-translations
which the unitary representation $\widetilde{U}$ of $\widetilde{\mathfrak{P}}{}^\uparrow_+$
implements on the Hilbert space $\cH$ are interpreted also as dynamical
evolutions of the system, and that these dynamical evolutions be stable in the sense
that their corresponding energy is always non-negative and that there should be
a common state of lowest energy, the vacuum state. This state is ``void of stable
particles'' but, as we shall see later, not void of correlations, and these have actually
a rich structure.
 \par
It is the subtle interplay of dynamical stability in the form of the spectrum condition
together with locality (or spacelike anti-commutalivity in the case of quantum fields carrying
half-integer spin) which is responsible for this richness. The condition of cyclicity
is mainly made for mathematical convenience; it says that all state vectors of the theory
can be approximated by applying polynomials of all field operators on the vacuum. In case of
the presence of a vacuum vector, this property could be sharpened to irreducibility, i.e.\
that allready all observables can be approximated by polynomials in the field operators.
This is actually equivalent to clustering of vacuum expectation values \cite{StrWigh,BLOT,Jost}. 
However, in a more general situation where there is no vacuum state for 
all time-evolutions (time-shifts), but e.g., a thermal equibrium state, irreducibility doesn't hold
in general.
 \par
While the Wightman framework captures apparently many essential aspects of (observable) quantum
fields and is so far not in obvious conflict with experiences gained in constructive quantum
field theory, there are some points which lead one to trade this framework for a still more 
abstract approach. Let me try to illustrate some of these points. The first is of
a more technical nature: In handling the --- in general --- unbounded field operators
$\Phi(F)$, subtle domain questions come into play whose physical significance is often not
entirely clear. More seriously, it might happen that the field operators $\Phi(F)$ do not
correspond to directly observable quantities, and then it is doubtful why they should be
regarded as the basic objects of the formal description of a physical theory, at least from
an operational point of view. Somehow related to this shortcoming, the $\Phi(F)$ aren't invariants
of the experimentally accessible quantities in the following sense: In general, one can find
for a given Wightman field $F \mapsto \Phi(F)$ other Wightman fields $F \mapsto
\tilde{\Phi}(F)$, subject to different field equations and commutation relations,
which yield the same $S$-matrix as the field $F \mapsto \Phi(F)$ (\cite{BorcherSmatrix},
see also \cite{Rehren} for a more recent instance of this fact).
 Apart from that, gauge-carrying quantum fields do not fit completely into the framework.
Assuming them to be local fields in the same sense as described above often leads to difficulties
with Hilbert space positivity, as e.g.\ in quantizing free electrodynamics.
This difficulty can be cured symptomatically by allowing $\cH$ to carry an inner product
that is not positive definite \cite{BLOT}. However, such a complication
makes technical issues, such as domain questions, even much worse.
 \par
Hence, there is considerable motivation to base the description of a relativistic quantum
system on observable quantities and to abandon the mainly classically inspired concept of
(a quantized version of) a field. In the case that $F \mapsto \Phi(F)$ is an observable
quantum field, one can pass to a description of this system which emphasizes the localization
of observables in space and time rather than their arrangement into ``field strengths'':
One can form, for each open subset $O$ of Minkowski spacetime $M \equiv \bR^4$, a $*$-algebra
of bounded operators
\begin{eqnarray} \label{gen}
 \cR(O) & = & \{ *\text{-algebra generated by all}\  A = f(\Phi(F))\,, \nonumber \\
        & & {} \quad f: \bR \to \bR \ \text{bounded}\,, \ \ \ 
             F = \overline{F} \ \text{has support in}\ O \}
\end{eqnarray}
The properties that one finds for the family of $*$-algebras $\cR(O)$, $O$ ranging
over the bounded subsets of $\bR^4$, form the conditions of the operator algebraic approach to
general quantum field theories according to Haag and Kastler \cite{HK,Haag}. These conditions
read as follows.
\begin{itemize}
\item[{\bf a)}] {\bf  Isotony:}
${}$\quad\ \ \ \ \ $ O_1 \subset O_2 \quad \Rightarrow \quad \cR(O_1) \subset \cR(O_2)$\\[4pt]
\item[{\bf b)}] {\bf  Covariance:} \ \
 $A \in \cR(O) \Leftrightarrow U(L)AU(L)^* \in \cR(L(O))$, \\[4pt]
${}$ \hspace*{3.2cm} or $U(L)\cR(O)U(L)^* = \cR(L(O))$
\item[{\bf c)}] {\bf  Locality:} If the space-time regions $O_1$ and $O_2$ 
are {\it causally separated}, then the corresponding operator algebras 
$\cR(O_1)$ and $\cR(O_2)$ {\it commute elementwise}:
$$ A \in \cR(O_1)\,,\ \ B \in \cR(O_2)\ \ \ \Rightarrow\ \ \ [A,B] = 0 $$
\item[{\bf d)}] {\bf  Spectrum condition and existence of the vacuum:}\\ As before in {\bf v)} and {\bf vi)}
\item[{\bf e)}] {\bf  Cyclicity of the vacuum:}\\[4pt] $\{ A\Omega : A \in \bigcup_O\cR(O)\}$
is dense in $\cH$
\item[{\bf f)}] {\bf Weak additivity}: If $\bigcup_i O_i$ contains $O$, then the
algebra generated by the $\cR(O_i)$ contains $\cR(O)$
\end{itemize}
We should emphasize that, adopting this framework as basis for a description of
a special relativistic quantum system, the crucial structural ingredient is the
assignment of not just a single operator algebra to the system but of operator
algebras $\cR(O)$ to the individual sub-regions $O$ of Minkowski spacetime. Each
$\cR(O)$ is generated by the observables which can be measured at times and locations
in $O$, and therefore one refers to the observables in $\cR(O)$ as those {\it localized in} $O$,
and to the $\cR(O)$ as {\it local observable algebras}. 
If actually there is a quantum field $F \mapsto \Phi(F)$ generating the local observable
algebras as in \eqref{gen}, then one may view it as a ``coordinatization''
of the family $\{\cR(O)\}_{O \subset M}$, the latter being the ``invariant'' object,
in analogy to a manifold built up from coordinate systems.
  \par
 A set of data
 $(\{\cR(O)\}_{O \subset M},U,\Omega)$ fulfilling the conditions just listed is called a
{\it quantum field theory in vacuum representation}.
One can consider other representations of a quantum field theory, e.g.\ thermal
representations, where the spectrum condition imposed on $U$ and the vacuum vector
condition imposed on $\O$ are replaced by the condition that the state
$\langle \O|\,.\,|\O\rangle$ be a thermal equilibrium state. We will encounter such
a situation later.
 \par
The reader might wonder at this point how charge carrying quantum fields fit into this
operator algebraic version of quantum field theory where up to now only observable quantities
have been mentioned. The answer is that charge carrying field operators arise in connection
with yet other Hilbert space representations of the quantum field field theory, i.e., of
the family of operator algebras $\{\cR(O)\}_{O \subset M}$. States in these 
Hilbert space representations cannot be coherently superposed with any state in the
vacuum representation. These charged representations are therefore called superselection
sectors. The analysis of superselection sectors and the full reconstruction of a compact
gauge group and of associated charge carrying quantum field operators from the structure
of superselection sectors can be regarded as being one of the greatest achievements in
axiomatic quantum field theory so far, but we shall not pause to explore these matters
and refer the reader to \cite{Haag,DR,Roberts-more} for further information.
 \par
It should be pointed out that all quantum fields obeying linear equations of motion
provide examples for the operator algebraic framework, by the relation \eqref{gen} [for
integer spin fields; for half-integer spin fields, one must instead define $\cR(O)$
by first constructing suitable bilinear expressions in the fields]. Moreover, there are
examples of interacting quantum fields in 2 and 3 spacetime dimensions and these are compatible
with the operator algebraic framework via \eqref{gen}.
 \par
The interplay between the spectrum conditon and locality puts non-trivial constraints
on quantum field theories and leads to interesting general results about their structure.
Prime examples are the PCT theorem, the spin-statistics relation (cf.\ \cite{StrWigh,BLOT,Haag}) and geometric
modular action. About the latter, perhaps less familiar, but highly fascinating issue
we have more to report in the following section.
\section{The Reeh-Schlieder Theorem and Geometric\\ Modular Action}
\setcounter{equation}{0}
In 1961, Helmut Reeh and Siegfried Schlieder showed that the conditions for a
quantum field theory of Wightman type, given above, lead to a remarkable consequence
\cite{ReehSchl}. Namely, let $O$ be any non-void open region in Minkowski spacetime,
and denote by $\cP(O)$ the $*$-algebra generated by all quantum field operators
$\Phi(F)$ where the test-functions are supported in $O$. Then the set of vectors
$\cP(O)\O$, $\O$ denoting the vacuum vector, is dense in the Hilbert space $\cH$. 
In other words, given an arbitrary vector $\psi \in \cH$, and $\epsilon > 0$,
there is a polynomial
\begin{equation} \label{twostar}
 \lambda_0 {\bf 1} + \sum_{j,k_j \le N} \Phi(F_{1,j}) \cdots \Phi(F_{k_j,j})
\end{equation}
in the field operatos, with $\lambda_0 \in \bC$ and $F_{\ell,j} \in C_0^\infty(O)$,
such that
\begin{equation} \label{threestar}
 ||\, \psi - ( \lambda_0 {\bf 1} + \sum_{j,k_j \le N} \Phi(F_{1,j}) \cdots \Phi(F_{k_j,j})) \O \,|| < \epsilon\,.
\end{equation}
In the operator algebraic setting of Haag and Kastler, the analogous property states
that the set of vectors $\cR(O)\O = \{ A\O : A \in \cR(O)\}$ is dense in
$\cH$ whenever $O$ is a non-void open set in $M$; equivalently, given $\psi \in \cH$
and $\epsilon > 0$, there is some $A \in \cR(O)$ fulfilling
\begin{equation} \label{star}
 ||\, \psi - A\O\,|| < \epsilon\,.
\end{equation}
This result by Reeh and Schlieder appears entirely counter-intuitive since it says that
every state of the theory can be approximated to arbitrary precision by acting
with operators (operations) localized in any arbitrarily given spacetime region
on the vacuum. To state it in a rather more drastic an provocative way
(which I learned from Reinhard Werner): By acting on the vacuum with suitable operations in a terrestial laboratory, an experimenter can create the Taj Mahal on (or even behind) the Moon!
 \par
One might thus be truely concerned that this unususal behaviour of relativistic 
quantum field theory potentially entails superluminal signalling. However,
despite the fact that such propositions have been made, this is not the case
(see \cite{Schlieder,HellwigKraus,BuchhYng} for some clarifying discussions).
We will also turn to aspects of this below in Sec.\ 4. A crucial point is that the 
operator $A = A_\epsilon$ in \eqref{star} depends on $\epsilon$ (and likewise,
the polynomial \eqref{twostar} in \eqref{threestar} depends on $\epsilon$), and
while $||A_\epsilon \O||$ will be bounded (in fact, close to 1) for arbitrarily small
$\epsilon$ (as follows from \eqref{star}), it will in general (in particular, with our
drastic Taj Mahal illustration) be the case that $A_\epsilon$ doesn't stay
bounded as $\epsilon \to 0$, in other words, $||A_\epsilon||$ diverges as
$\epsilon$ tends to $0$.
 \par
In keeping with the standard operational interpretation of quantum theory \cite{Kraus},
$||A_\epsilon||/||A_\epsilon\O||$ is to be viewed as the ratio of cost vs.\ effect
in the attempt to create a given state (Taj Mahal on the Moon) by local operations
(in a laboratory on Earth, say) \cite{Haag}. In other words, upon testing for
coincidence with the ``Taj Mahal state $\psi$'', it takes on average an ensemble
of $||A_\epsilon||/||A_\epsilon\O||$ samples failing in the coincidence test to find a 
single successful coincidence. And in our illustration, the ratio
$||A_\epsilon||/||A_\epsilon\O||$ will be an enormous number. A rough estimate
can be based on the decay of vacuum correlations in quantum field theory. The
order of magnitude of that decay is approximately given by ${\rm e}^{-d/\l_c}$, where
$d$ denotes the spatial distance of the correlations and $\l_c$ is the Compton
wave length of the stable particles under consideration; then $1/{\rm e}^{-d/\l_c}$
is a rough measure for $||A_\epsilon||/||A_\epsilon\O||$ (when $\epsilon$ is very
small compared to 1). Taking for instance electrons as stable particles,
 and the distance
Earth-Moon for $d$, one obtains an order of magnitude of about $10^{-10^{20}}$ for
${\rm e}^{-d/\l_c}$. This shows that one can hardly construe a contradiction to special
relativity on account of the Reeh-Schlieder theorem.
 \par
Nevertheless, for distances that are comparable to the Compton wavelength, the Reeh-Schlieder-theorem does predict a behaviour of the correlations in the vacuum state which 
is in principle experimentally testable, and is of truely quantum nature in the sense
that they entail quantum entanglement over subsysems, as will be seen later in
Sec.\ 4.
 \par 
We will complement the previous discussion by a couple of remarks. 
\\[6pt]
(3.4) The mathematical cause for the Reeh-Schlieder theorem lies
in the spectrum condition, which entails that, for each $\psi$ in the Hilbert
space of a quantum field theory's vacuum representation, the function
\begin{equation} \label{anlbound}
(a_1,\ldots,a_n) \mapsto \langle \psi,U(a_1)A_1U(a_2)A_2 \cdots U(a_n)A_n\O \rangle\,,
\ \ A_j \in \cR(O)\,,  \ a_j \in \bR^4\,,
\end{equation}
is the continuous boundary value of a function which is analytic in a conical
subregion of $\bC^{4n}$. Hence, if the expression \eqref{anlbound} vanishes when the $a_j$ are
in an arbitrarily small open subset of $\bR^4$, then it vanishes for all $a_j \in \bR^4$.
Together with weak additivity one can conclude from this that any vector $\psi$ which
is orthogonal to $\cR(O)\O$ is actually orthogonal to $\bigcup_O \cR(O)\O$ and hence,
by cyclicity of the vacuum vector, $\psi$ must be equal to $0$.
\\[6pt]
(3.5) There are many other state vectors $\xi \in \cH$ besides the
vacuum vector for which the Reeh-Schlieder theorem holds as well, i.e.\ for which
$\cR(O)\xi = \{ A \xi: A \in \cR(O)\}$ is a dense subset of $\cH$ whenever $O \subset M$
is open and non-void. In fact, one can show that there is a dense subset $\cX$ of $\cH$ so
that every $\xi \in \cX$ has the property that $\cR(O)\xi$ is dense in $\cH$ as soon
as $O \subset M$ is a non-void open set \cite{DixMar}. Now, every element $\xi \in \cX$
(assumed to be normalized) induces a state (expectation value functional)
$$ \o_{\xi}(A) = \langle \xi,A\xi \rangle\,, \quad A \in \cR(\bR^4)\,, $$
and owing to the Reeh-Schlieder property of the vectors $\xi \in \cX$, $\o_{\xi}$
will have long-range correlations, meaning that generically
$$ \o_{\xi}(AB) \ne \o_{\xi}(A)\o_{\xi}(B) $$
for $A \in \cR(O_{\rm A})$ and $B \in \cR(O_{\rm B})$ even if the spacetime
regions $O_{\rm A}$ and $O_{\rm B}$ are separated by an arbitrarily large spacelike distance.
However, even though the set of vectors $\xi$ inducing such long-lange correlations 
is dense in the
set of all state vectors in $\cH$, there are in general also very many uncorrelated states.
In fact, under very general conditions it could be shown that, as soon as a pair of
(finitely extended) spacetime regions $O_{\rm A}$ and $O_{\rm B}$  separated by a
non-zero spacelike distance is given, together with a pair of vectors $\xi_{\rm A}$ and
$\xi_{\rm B}$ in $\cH$ inducing states $\o_{\xi_{\rm A}}$ and $\o_{\xi_{\rm B}}$ on the
local observable algebras $\cR(O_{\rm A})$ and $\cR(O_{\rm B})$, respectively, there is a
state vector $\eta \in \cH$ inducing a state $\o_\eta$ on $\cR(\bR^4)$ with the property
$$ \o_{\eta}(AB) = \o_{\xi_{\rm A}}(A)\o_{\xi_{\rm B}}(B)\,, \quad A \in \cR(O_{\rm A}),\ \ B \in
\cR(O_{\rm B})\,.$$
That is to say, in restriction to the algebra of observables associated to the region
$O_{\rm A} \cup O_{\rm B}$ the state $\o_{\eta}$ coincides with the (prescribed)
product state induced by the pair of states $\o_{\xi_{\rm A}}$ and $\o_{\xi_{\rm B}}$ which
has no correlations between the subsystems $\cR(O_{\rm A})$ and $\cR(O_{\rm B})$. We should like
to refer the reader to \cite{BuWi,SJS-RMP2} for considerable discussion on this issue.
\\[6pt]
(3.6) There are states $\xi \in \cX$ for which the Reeh-Schlieder correlations
are much stronger that in the vacuum $\O$, and in such states the correlations
are sufficiently strong so that they can be used for quantum teleportation over macroscopic
distances as has been demonstrated experimentally \cite{ZeilingerundCo}. While this is
perhaps intuitively less surprising than for the case of the vacuum state since the states $\xi$
have some  ``material content'' to which one could ascribe the storage of correlation
information, it should be kept in mind that also here the correlations are non-classical, i.e.\
they manifestly exemplify quantum entanglement.
\\[6pt]
(3.7) In the Haag-Kastler setting, local commutativity and the Reeh-Schlieder theorem
together imply that any local operator $A \in \cR(O)$, $O$ open and bounded, which annihilates the
vacuum: $A\O = 0$, must in fact be equal to the zero operator, $A =0$.
As a consequence, for the vacuum vector $\O$ (as well as for any other $\xi \in \cX$ having
the Reeh-Schlieder property) it holds that 
$$ \langle \O,A^*A \O \rangle > 0 $$
for all $A \in \cR(O)$ with $A \ne 0$, $O$ open and bounded.
This may be interpreted as the generic presence of vacuum fluctuations; every {\it local} counting
instrument will give a non-zero expectation value in the vacuum state. This is, actually,
a situation where relativistic quantum field theory deviates from quantum mechanics.
(Quantum mechanics needs to postulate the existence of fluctuations as e.g.\ in the semiclassical
theory of radiation to account for spontaneous emission.)
 \par
A related mathematical argument shows that quantities like the energy density will fail
to be pointwise positive in the quantum field setting, in contrast to their classical
behaviour. Yet, the spectrum condition puts limitations to the failure of
positivity. For this circle of questions, we recommend that the reader consults the 
review article \cite{FewsterICMP}. 
\\[10pt]
Now, in order to turn to the discussion of ``geometric modular action'', we need to introduce
some notation. We consider a generic {\it von Neumann algebra} $\cR$ acting on a Hilbert space
$\cH$, together with a unit  vector $\O \in \cH$ which is assumed to be {\it cyclic and separating}
for $\cR$. To explain the terminology, $\cR$ is a von Neumann algebra acting on $\cH$ if
$\cR$ is a weakly closed (in the sense of convergence of expectation values) $*$-subalgebra
of $B(\cH)$ containing the unit operator. One can show that this is equivalent to the property that
$\cR$ coincides with its double commutant $\cR''$, where the commutant $\cC'$ of a subset $\cC$ of
$B(\cH)$ is defined as $\cC' = \{B \in B(\cH) : BC = CB \ \forall\ C \in \cC\}$, and the
double commutant is then defined by $\cC'' =(\cC')'$. One says that $\O \in \cH$ is cyclic
for $\cR$ if $\cR\O$ is dense in $\cH$ --- in view of our previous discussion, this is the
same as saying that the Reeh-Schlieder property holds for $\O$, with respect to the algebra $\cR$.
Moreover, one says that $\O$ is separating for $\cR$ if $A \in \cR$ and $A\O = 0$ imply
$A =0$, and this is equivalent to $\langle \O,A^*A \O\rangle > 0$ for all $A \in \cR$ different
from $0$. One can in fact show that $\O$ is cyclic for $\cR$ if and only if $\O$ is separating
for $\cR'$, and vice versa.
 \par
Given a von Neumann algebra $\cR$ on a Hilbert space $\cH$ and a cyclic and separating unit
vector, $\O \in \cH$, for $\cR$, there is a canonical anti-linear operator 
$$ S: \cR\O \to \cR\O\,, \quad A\O \mapsto S(A\O) :=A^*\O $$
associated with these data. By cyclicity of $\O$ for $\cR$, the set $\cR\O = \{A\O:
A \in \cR\}$ is a dense linear subspace of $\cH$, so the operator is densely defined;
furthermore, to assign the value $A^*\O$ to the vector $A\O$ in the domain of $S$ is a
well-defined procedure in view of the assumption that $\O$ is separating for $\cR$.
The anti-linearity of $S$ is then fairly obvious. What is less obvious is the circumstance
that the operator $S$ is usually unbounded (provided $\cH$ is infinite-dimensional).
Nevertheless, one can show that $S$ is a closable operator and thus the closure of $S$
(which we denote here again by $S$) possesses a polar decomposition, i.e.\ there is
a unique pair of operators $J$ and $\D$ so that $S$ can be written as
$$ S = J \D^{1/2} $$
and where $J: \cH \to \cH$ is anti-linear and fulfills $J^2 = {\bf 1}$ while $\D = S^*S$
is positive (and selfadjoint on a suitable domain, and usually unbounded). This is nothing
but the usual polar decompositon of a closable operator, with the slight complication
that the operator $S$ is, by definition, anti-linear instead of linear. See, e.g.,
\cite{BratRob1} for further information. 
 \par
The operators $J$ and $\D$ are called the {\it modular conjugation}, and 
{\it modular operator}, respectively, corresponding to the pair $\cR,\O$. Often,
$J$ and $\D$ are also referred to as the {\it modular objects} of $\cR,\O$. Their
properties have been investigated by the mathematicians Tomita and Takesaki and hence they
appear also under the name {\it Tomita-Takesaki modular objects}.
The important properties of $J$ and $\D$ which were discovered by Tomita and
Takesaki (see, e.g., \cite{Takesaki,BratRob1,Borchers} for a full survey of
the mathematical statements which we make in what follows) are, first, that 
the adjoint action of $J$ maps 
$\cR$ onto its commutant $\cR'$: $A \in \cR \Leftrightarrow JAJ \in \cR'$. This is written in shorter notation as $J\cR J = \cR'$. One also has that $J\O = \O$. Secondly,
since $\D$ is an invertible non-negative selfadjoint operator, ${\rm ln}(\D)$
can be defined as a selfadjoint operator by the functional calculus, and hence one can
define a one-parametric unitary group $\D^{it} = {\rm e}^{it{\rm ln}(\D)}$, $t \in \bR$, on $\cH$, called the {\it modular group} of $\cR$ and $\O$. 
It has the property that its adjoint action leaves $\cR$ invariant, i.e.\
$A \in \cR \Leftrightarrow \D^{it} A \D^{-it} \in \cR$, or simply
$\D^{it} \cR \D^{-it} = \cR$. Moreover, $\D^{it}\O = \O$ holds for all $t \in \bR$.
A third property relates to the spectral behaviour of the unitary group $\{\D^{it}\}_{t \in \bR}$. Namely, the state $\o_{\O}(A) = \langle \O,A\O\rangle$ on $\cR$
fulfills the KMS (Kubo-Martin-Schwinger) boundary condition with respect to the adjoint
action of $\D^{it}$, $t\in \bR$, at inverse temperatur $\b = 1$.
 \par
Let us explain the terminology used here. If $\cR$ is a von Neumann algebra modelling
the observables of a quantum system and $\{\sigma_t\}_{t\in\bR}$ is a one-parametric
(continuous) group of automorphisms of $\cR$ modelling the dynamical evolution
of the system, then a density matrix state $\o_{\rho}(A) = {\rm trace}(\rho A)$ on
$\cR$ is said to fulfil the KMS boundary condition with respect to $\{\sigma_t\}_{t\in\bR}$ (shorter: is a KMS state for $\{\sigma_t\}_{t\in\bR}$) at inverse
temperatur $\b > 0$ provided that the following holds: Given any pair of elements
$A,B \in \cR$, there exists a function $F_{AB}$ which is analytic on the complex
strip $\cS_{\b} =\{ t + i\eta: t \in \bR\,,\ 0 < \eta < \beta \}$, and is
 continuous on the closure of the strip $\cS_{\b}$, with the boundary
values
$$ F_{AB}(t) = \o_{\rho}(\sigma_t(A)B)\,, \quad F_{AB}(t + i\beta) = \o_{\rho}(B\sigma_t(A))\,, \quad \ \ t \in \bR\,.$$
For a quantum mechanical system with a Hamilton operator $H$ such that ${\rm e}^{-\b H}$
is a trace-class operator $(\b > 0)$, one can form the density matrices
$$\rho^{\b} = \frac{1}{{\rm trace}({\rm e}^{-\b H})}{\rm e}^{-\b H}$$ 
and one can show that the corresponding Gibbs states $\o_{\rho^{\b}}$ are KMS states at inverse temperature $\b$ for the dynamical
evolution given by $\sigma_t(A) = {\rm e}^{itH} A {\rm e}^{-itH}$\,. Haag, Hugenholtz
and Winnink \cite{HHW} have shown that states of an infinite quantum system --- being
modelled by $\cR$ and  $\{\sigma_t\}_{t\in\bR}$ --- which are suitably approximated by
Gibbs states of finite subsystems, are under very general conditions also KMS states, and thus
the KMS boundary condition is viewed as being characteristic of thermal equilibrium states.
 \par
Therefore, if $\o_{\O}$ is a KMS state with respect to the (adjoint action of the)
modular group $\{\D^{it}\}_{t\in \bR}$ of $\cR$,$\O$, this signalizes that there is
some relation to physics provided that $\{\D^{it}\}_{t\in\bR}$ can be interpreted
as dynamical evolution of a quantum system. This is not always the case, but the converse
always holds true: Suppose that a quantum system dynamical system consisting of $\cR$ and
$\{\sigma_t\}_{t\in\bR}$ and a KMS state $\o_{\rho}$ at inverse temperature $\b > 0$ is given.
Then one can pass to the {\it GNS (Gelfand-Naimark-Segal) representation} associated with
$\cR$ and $\omega_\rho$. This is a triple $(\pi^{\rho},\cH^{\rho},\O^{\rho})$ where $\cH^{\rho}$
is a Hilbert space, $\pi^{\rho}$ is a representation of $\cR$ by bounded linear operators on 
$\cH^{\rho}$ (which may differ from the ``defining'' representation of $\cR$
that is pre-given since the
elements of $\cR$ act as bounded linear operators on a Hilbert space $\cH$) and $\O^{\rho}$
is a unit vector in $\cH^{\rho}$ which is cyclic for $\pi^{\rho}(\cR)$ and wtih
$\o^{\rho}(A) = \langle \O^{\rho},\pi^{\rho}(A) \O^{\rho} \rangle$. In this GNS representation,
$\{\sigma_t\}_{t\in\bR}$ is implemented by the (rescaled) modular group $\{\D^{it/\b}\}_{t \in \bR}$
corresponding to $\pi^{\rho}(\cR)''$ and $\O^{\rho}$: $\pi^{\rho}(\sigma_t(A)) =\D^{it/\b} \pi^{\rho}(A)
\D^{-it/\b}$.
 \par
Tomita-Takesaki theory has had a considerable impact on the development of
operator algebra theory. Owing to its relation to thermal equilibrium states, it
has also found applications in quantum statistical mechanics. It took longer,
however, until a connection between Tomita-Takesaki modular objects and the
action of the Poincar\'e group was revealed in the context of relativistic quantum
field theory. Such a connection was established in the seminal work of 
Bisognano and Wichmann \cite{BiWi}. To explain their result, let 
$(x^0,x^1,x^2,x^3)$ denote the coordinates of points in Minkowski spacetime in some
Lorentzian frame. Then let $W = \{x =(x^0,x^1,x^2,x^3) \in \bR^4 :
x^1 > 0,\ - x^1 < x^0 < x^1\}$ denote the ``right wedge region'' with respect
to the chosen coordinates.
Moreover, we shall introduce the following maps of Minkowski spacetime:
$$ {\bf j}: (x^0,x^1,x^2,x^3) \mapsto (-x^0,-x^1,x^2,x^3) $$
which is a reflection about the spatial $x^2$-$x^3$ plane together with
a time-reflection, and 
$$ \Lambda_1(\theta) = \left( \begin{array}{cccc}
                              \text{cosh}(\theta) & - \text{sinh}(\theta) & 0 & 0 \\
                               -\text{sinh}(\theta) & \text{cosh}(\theta) & 0 & 0\\
                                     0 & 0 & 1 & 0\\
                                      0 & 0 & 0 & 1
                                  \end{array} \right)\,, \quad \theta \in \bR\,,$$
the Lorentz boosts along the $x^1$-axis, which map $W$ onto itself.
 \par
Now consider a quantum field theory of the Haag-Kastler type (in vacuum
representation), where it is also assumed
that the local algebras of observables $\cR(O)$ are generated by a Wightman-type
quantum field $F \mapsto\Phi(F)$ as in \eqref{gen}. It will also be assumed that the $\cR(O)$ are
actually von Neumann algebras, so that one has $\cR(O) = \cR(O)''$ for open, bounded
regions. Then one can also built the algebra of observables located in the wedge region
$W$,
$$ \cR(W) = \{ A \in \cR(O) : O \subset W\}''\,.$$
We will denote by $J$ the modular conjugation and by $\{\D^{it}\}_{t\in\bR}$ the
modular group, respectively, associated with $\cR(W)$ and the vacuum vector $\O$.
These are well-defined since the vacuum vector is, by the Reeh-Schlieder theorem,
cyclic and separating for $\cR(W)$.
With these assumptions, Bisognano and Wichmann \cite{BiWi} found the following remarkable result.
 \\[6pt]
(3.8) {\bf Theorem} \\
The following relations hold:
\begin{eqnarray*}
 \Delta^{it} & = & U(\Lambda_1(2\pi t)) \\
 J \cR(O) J & = & R( {\bf j}(O)) \,,\ \ \ \text{moreover},\\
 J \Phi(F) J & = & \Phi(\overline{F \circ {\bf j}})\,,\\
 J U(L) J & =& U( {\bf j} \circ L \circ {\bf j}) \,, \quad L \in \mathfrak{P}^\uparrow_+
\end{eqnarray*}
Here, $U$ denotes the unitary representation of the Poincar\'e group belonging to
the quantum field theory under consideration, and we have written
$U(\Lambda_1(2\pi t))$ for the unitary representation of the Lorentz boost $\Lambda_1(2\pi t)$.
 \par
The remarkable point is that by this theorem, the modular conjugation and modular
group associated with $\cR(W)$ and $\O$ aquire a clear-cut geometric meaning. Moreover,
since the adjoint action of $J$ involves, in its geometric meaning, a time and
space reflection, it induces a PCT symmetry in the following way:\\[6pt]
The rotation $D_{(2,3)}$ by $\pi = 180^\circ$ in the $(x^2,x^3)$ plane is contained in
the proper, orthocronous Poincar\'e group, and 
$$ {\bf j} \circ D_{(2,3)} = D_{(2,3)} \circ {\bf j} = PT : x \mapsto -x $$
is the total inversion.\\[6pt]
Then $\Theta = J U(D_{(2,3)})$ is a {\it PCT operator}:
$\Theta$ is anti-unitary and fulfills $\Theta^2 = \eins$, and
\begin{eqnarray*}
\Theta \Omega & = & \Omega\\
\Theta \cR(O) \Theta & = & \cR(PT(O)) \\
\Theta \Phi(F) \Theta & = & \Phi(\overline{F \circ PT})\\
\Theta U(L) \Theta & = & U(PT \circ L \circ PT)\,.
\end{eqnarray*}
 Because of the geometric significance of the modular objects $J$ and $\{\D^{it}\}_{t\in\bR}$ one also says that the Bisognano-Wichmann theorem is an instance of ``geometric
modular action'' (although this term is actually used also in a wider context).
The concept of ``geometric modular action'' has been used quite fuitfully in
the analysis of general quantum field theories over the past years and has led
to remarkable progress and insights. We cannot get into this matter in any depth
and instead we refer the reader to the comprehensive review by Borchers \cite{Borchers};
we will only comment on a few aspects of geometric modular action by way of a couple
of remarks.
\\[6pt]
(3.9)  Because of $\Delta^{it} = U(\Lambda_1(2 \pi t))$, the vacuum state
functional $\langle \Omega ,\,.\,\Omega \rangle$ restricted to 
$\cR(W)$ is a KMS state, i.e.\ a  thermal equilibrium state.
More precisely, an observer following the trajectory
$$ \gamma_a(t) = \Lambda_1(t)\left( \begin{array}{c} 0 \\ 1/a \\ 0 \\ 0 \end{array} \right)$$
will register the (restriction of the) vacuum state along his or her trajectory as
a thermal equilibrium state at absolute temperature
$$  T_a = \frac{\hbar a}{2\pi k c}\,, $$
where here we have explicitly inserted $\hbar$, Boltzmann's constant $k$ and the
velocity of light $c$.
This is called the {\it Fulling-Unruh-effect} \cite{Ful,Unr}.
It has been noted by Sewell \cite{Sew,SumVer} that a similar form of geometric modular action for 
quantum fields on the 
Schwarzschild-Kruskal spacetime can be viewed as a variant of the Hawking effect. 
\\[6pt]
(3.10) The relation of Tomita-Takesaki objects to the action of the Poincar\'e
group which is displayed by the Bisognano-Wichmann theorem is only realized if the
observable algebras with respect to which the Tomita-Takesaki objects those belonging to
wedge regions --- i.e.\ any Poincar\'e-transform of $W$. For observable algebras
$\cR(O)$ belonging to bounded regions, the corresponding modular objects have in general
 no clear geometric
meaning. An exception is the case of a conformal quantum field theory when $O$
is a double cone (see \cite{Borchers}
and literature cited there). \\[6pt]
(3.11)
If $a$ is a lightlike vector parallel to the future lightlike boundary
of $W$, let
$$ J_a = \text{modular conjugation of}\ \ \cR(W +a),\Omega $$
Then one can show that
$$ U(-2a) = J_0J_a \,,$$
i.e.\ the modular conjugations encode the translation group --- together with
the spectrum condition. Since the modular group of $\cR(W)$
induces the boosts, it appears that the complete unitary action of the
Poincar\'e group can be retrieved from the modular objects of observable algebras
belonging to a couple of wedge regions in suitable position to each other,
with a common vacuum vector.  And indeed, a careful analysis has shown that
this is possible under general conditions \cite{KahlWies,Borchers}. This opens the possibility
to approach the problem of constructing (interacting) quantum field theories in
a completely novel manner, where one starts with a couple of von Neumann algebras
together with a common cyclic and separating vector, and where the associated
modular objects fulfil suitable relations so that they induce a representation of
the Poincar\'e group. See \cite{SchroerWies,BuLechner} for perspectives, first steps and results
around this circle of ideas.
\\[6pt]
(3.12) It should also be pointed out that geometric modular action can be 
understood in a more general sense than above where the modular objects associated
with the vacuum and algebras of observables located in wedge-regions induce point-transformations
on the manifold --- in our present discussion, always Minkowski spacetime --- on which
the quantum field theory under consideration lives.  A more general criterion of geometric
modular action would, e.g., be the following: Given a family of observable
(von Neumann) algebras  $\{R(O)\}_{O \subset M}$ indexed by the open (and bounded)
subsets of a spacetime manifold $M$, and a vector $\O$ in the Hilbert space representation
of that family, one can try  to find a sub-family $\{R(\tilde{O})\}_{\tilde{O} \in \tilde{K}}$
(where $\tilde{K}$ is a collection of subsets of $M$, sufficiently large so that 
a base of the topology of $M$ can be generated by countable intersections and
unions of members in $\tilde{K}$, say) with the property that the adjoint action of
the modular conjugation $J_{\hat{O}}$ of $\cR(\hat{O}),\O$, where $\hat{O}$ is
any element of $\tilde{K}$, maps the family  $\{R(\tilde{O})\}_{\tilde{O} \in \tilde{K}}$
onto itself. This would be a generalized form of geometric modular action.
In the light of the Bisognano-Wichmann theorem, for the case of Minkowski
spacetime one would take the collection of wedge regions as $\tilde{K}$
and the vacuum vector as $\O$. But there
are instances where precisely such a generalized form of geometric modular action is
realized when taking for $M$ e.g.\ Robertson-Walker spacetimes. For more discussion
on this intriguing generalization of geometric modular action, see \cite{BuMuSu}.  
\section{Relativistic Quantum Information Theory: Distillability in
Quantum Field Theory}
\setcounter{equation}{0}
The final section of this contribution is devoted to a subject which
seems to be of growing interest nowadays \cite{BeckGottNielPres,EggeSchlWer,PerTer,RRS}: The attempt to bring together
the flourishing discipline of quantum information theory with the principles
of special relativity. Since quantum information theory is based on the principles of
quantum mechanics and since quantum field theory is the theory which unifies
quantum mechanics and special relativity, it appears entirely natural to discuss
issues of relativistic quantum information theory in the setting of quantum field
theory.
 \par
There are, of course, several foundational issues one might wish to discuss when
studying a prospective merging of quantum information theory and special relativity
even in the established setting of quantum field theory. One of them might be the
so far omitted discussion on quantum measurement theory within quantum field theory.
In view of the Reeh-Schlieder theorem, one may suspect delicate problems at this 
point --- in fact, there are numerous discussions on the nature of locality/nonlocality
in quantum (information) theory, where sometimes the various 
authors don't agree on precisely what sort
of locality is attributed to which object or structure within a particular theoretical
framework. Our approach here is operational, and we refer to works already cited \cite{Schlieder,HellwigKraus} for some discussion on measurement in quantum field theory.
 \par
This said, we limit ourselves here to studying a very particular concept which
has been developed and investigated in non-relativistic quantum information theory in
the context of relativistic quantum field theory: The concept of
{\it distillability} of quantum states. Very roughly speaking, one can say that
distillable quantum states contain ``useful'' entanglement that can be enhanced,
at least theoretically, so that it can be used as a resource for typical 
telecommunication tasks such as quantum cryptography or quantum teleportation
\cite{?c,Ekert,ZeilingerundCo}. (For a more detailed exposition of the formal apparatus
of quantum information theory and important references, we recommend the
review by M.\ Keyl \cite{Keyl}.)
To make this more precise, we will now have to specify our setting at a more formal level.
Everything what follows is taken from a joint publication with R.\ Werner \cite{VerWer}.
 \par
First, we will say that a {\it bipartite system} is a pair of mutually commuting
$*$-subalgebras $\cA$, $\cB$ of $B(\cH)$ for some Hibert space $\cH$. Usually, we
will in fact assume that both $\cA$ and $\cB$ are von Neumann algebras; one could also
generalize the setting by only requiring that $\cA$ and $\cB$ are $*$-subalgebras
of a common $C^*$-algebra.
 \par
In the quantum field theoretical context, $\cA$ will be identified with 
$\cR(O_{\rm A})$ and $\cB$ with $\cR(O_{\rm B})$ for a pair of (bounded)
spacetime regions $O_{\rm A}$ and $O_{\rm B}$ which are causally separated.
Quite generally, $\cA$ represents the algebra of observables in a laboratory
controlled by a physicist named `Alice' and $\cB$ represents the algebra of
observables in a laboratory controlled by another physicist called `Bob'.
The prototypical example of a bipartite system
in (non-relativistic) quantum information theory is the situation where 
$\cH = \cH_{\rm A} \otimes \cH_{\rm B}$, and where $\cA = B(\cH_{\rm A}) \otimes {\bf 1}$
and where $\cB = {\bf 1} \otimes B(\cH_{\rm B})$. The situation in relativistic quantum
field theory can be a bit more complicated.
 \par
Let $\cA,\cB \subset B(\cH)$ form a bipartite quantum system, and let $\o(X) = \text{trace}(\rho X)$, for some density matrix $\rho$ on $\cH$, be a state on $B(\cH)$. We say that the
state $\o$ is a {\it product state} on the bipartite system if 
$\o(AB) = \o(A)\o(B)$ holds for all $A \in  \cA$ and all $B \in \cB$. Moreover,
$\o$ is called {\it separable} on the bipartite system if it is a limit (in the sense
of convergence of expectation values) of convex combinations of product states.
Then, $\o$ is called {\it entangled} on the bipartite system if it is not
separable.
 \par
Entanglement of states on bipartite systems is a typical quantum phenomenon with no
counterpart in classical physics. As is well known, the Einstein-Podolsky-Rosen paradoxon
really centers about entangled states, as has been clarified and formalized by John Bell
(see the reprint collection \cite{WheelerZurek} for the relevant references and comments,
and the textbook \cite{Peres:book} for a more modern and simpler discussion). As mentioned, nowadays
entanglement is viewed as a resource for tasks of quantum communication, and this
circumstance has motivated several studies on the ``degree'' or ``quality'' of entanglement
that a state may have (see, again, the review \cite{Keyl} for 
discussion and references). One possible measure of ``entanglement strength'' is provided by
the {\it Bell-CHSH correlation} \cite{CHSH,SumWer87a}. This is a number, $\b(\o)$, which
is assigned to any state $\o$ of a bipartite system  $\cA,\cB \subset B(\cH)$ as
$$ \b(\o) = \sup_{A,A',B,B'}\,\o(A(B' + B) + A'(B' - B)) $$
where the supremum is taken over all hermitean $A,A' \in \cA$ and $B,B' \in \cB$
whose operator norm is bounded by 1.
Separable states always have $\b(\o) \le 2$. This case is referred to by saying that
$\o$ {\it fulfills the Bell-CHSH inequalities}. States $\o$ for which $\b(\o) > 2$
are said to {\it violate the Bell-CHSH inequalities}; such states are entangled.
The maximal number which $\b(\o)$ can assume is $2\sqrt{2}$ \cite{Cirelson}, and states
for which $\b(\o) = 2\sqrt{2}$ are said to {\it violate the Bell-CHSH inequalities maximally}.
In a sense, one may view a state $\o_1$ more strongly entangled than a state $\o_2$ if
$\b(\o_1) > \b(\o_2)$.
 \par
Let us consider a particularly simple system where $\cH$ = $\bC^2 \otimes \bC^2$, with
$\cA = B(\bC^2) \otimes {\bf 1}$ and $\cB = {\bf 1} \otimes B(\bC^2)$, where
$B(\bC^2)$ is a perhaps slightly unusual way to denote the algebra of complex $2 \times 2$
matrices.    
A state violating the Bell-CHSH inequalities maximally is given by the singlet state
$\o_{\rm singlet}(X) = \langle \psi_{\rm singlet}, X \psi_{\rm singlet} \rangle$, $X \in
B(\bC^2 \otimes \bC^2)$, where
$$ \psi_{\rm singlet} = \frac{1}{\sqrt{2}}(|0\rangle \otimes |1\rangle - |1\rangle \otimes |0\rangle)\,;$$
here, $|0\rangle$ and $|1 \rangle$ denote the two orthonormalized eigenvectors of the
Pauli-matrix $\sigma_z$. There are, in fact, experimental situations in quantum optics where
the singlet state can be realized to a high degree of accuracy. In these situations,
one identifies $|0\rangle$ and $|1\rangle$ with the two orthonormal polarization states of 
photons  which are linearly
polarized with respect to chosen coordinates perpendicular to the direction
of propagation. One can prepare a source (state) producing an ensemble of 
 pairs of polarized photons
in the singlet state and send --- e.g.\ through optical fibres over long distances ---
one member of each ensemble pair to the laboratory of Alice (whose observables,
regarding the polarization of the photons, are
represented by $\cA$) and the other member of the same pair to the laboratory of Bob
(whose polarization observables are represented by $\cB$).
 In this way, Alice and Bob have 
access to a common entangled state $\o_{\rm singlet}$ which they may use for
carrying out tasks of quantum communication.  The singlet state (or rather, any
singlet-type state) is, in this sense, the
best suited state owing to its ``maximal'' entanglement which is reflected by its maximal
violation of the Bell-CHSH inequalities. Some experimental realizations
and applications can be found e.g.\ in \cite{ZeilingerundCo}.
 \par
There are entangled states $\o$ which are not as strongly entangled as $\o_{\rm singlet}$,
but contain still enough entanglement so that a sub-ensemble of photon pairs can
be ``distilled'' from $\o$ which coincides with $\o_{\rm singlet}$ to high accuracy and may then
be used for carrying out quantum communication tasks. 
To make such a ``distillability'' an attribute of the given state $\o$, one must
ensure that the 
distillation process only enhances the entanglement already present in the
given state $\o$, and doesn't induce previously non-existing entanglement.
One tries to capture this requirement by demanding that the process of distillation
involves only local operations and classical communication (LOCC) \cite{BDSW,Popescu94,Keyl}.
 \par
The idea behind LOCC is best illustrated by a simple example. We assume
that both Alice and Bob operate a two-valued instrument
in each of their laboratories. A two-valued instrument (i) takes an incoming state,
(ii) puts out either of two classical values (``readings''), say ``+'' or ``$-$''
and (iii) changes the state into a new output state depending on the values of the
classical readings, i.e.\ the values ``$+$'' or ``$-$''.
Thus, if the source (represented by the state $\o$) produces a pair of polarized
photons, then the pair member running to Alice passes her instrument while the  
other pair member travels to Bob and passes his instrument. The pair members are
then subjected to state changes --- operations --- taking place individually at the 
sites of the laboratories of Alice and Bob, respectively, and are thus local
(assuming that the operations are active at mutual spacelike separation); put differently,
Alice's instrument operates only on the pair member in her laboratory and likewise Bob's
instrument operates only on the pair member in his laboratory.
We further suppose that Alice and Bob agree to discard all photon pairs except those 
which on passing their instruments have yielded in both cases the ``$+$'' reading.
Since they don't now beforehand what the values of these readings will be, they have to
inform each other about the readings' values of their instruments {\it after} both members of
each photon pair have passed through. This requires ``two-way classical communication''
between Alice and Bob. Then, after a large number of photon pairs (corresponding, in idealization,
to the original ensemble of the state $\o$) has passed the instruments, and having discarded all the pairs not giving the ``$+$'' reading, Alice and Bob hold (in each lab, members of)
a smaller number (a subensemble) of photon pairs which have been subjected to local operations
mediated by the instruments. This new subensemble may correspond to a state with stronger
entanglement, and if, in this way, a subensemble can be produced which
approximates the singlet state $\o_{\rm singlet}$ to arbitrary precision, then the 
original state $\o$ is called {\it distillable}. Strictly speaking, we should
call the state {\it 1-distillable}, the qualifier ``1'' referring to only ``1 round'' of
instrument application and classical communication for each photon pair, since one can envisage more complicated schemes of using localized (multi-valued) instruments and classical communication between
Alice and Bob that are still in compliance with the idea of local operations and classical communication. But then, any state which is 1-distillable will also be distillable according
to a more general scheme, so that 1-distillability is in this sense the most stringent
criterion.
 \par
Now we need to give a mathematical description of 1-distillability of a state $\o$.
In the present simple case, the mathematical image of a two-valued instrument
in Alice's laboratory is given by two completely postive maps $T_{\pm} : \cA \to \cA$
with $T_+({\bf 1}) + T_-({\bf 1}) = {\bf 1}$. Likewise, in Bob's laboratory,
his two-valued instrument is given by a pair of completely positive maps 
$S_{\pm} : \cB \to \cB$ with $S_+({\bf 1}) + S_-({\bf 1}) = {\bf 1}$.
The subensemble that Alice and Bob select from the original state $\omega$ 
corresponds to the positive functional $\cA \otimes \cB \owns x \otimes y 
\mapsto \o(T_+(x)S_+(y))$,
which is turned into a state, $\cA \otimes \cB \owns
 x \otimes y \mapsto \o(T_+(x)S_+(y))/\o(T_+({\bf 1})S_+({\bf 1}))$,
 upon normalization. Let us denote this new state by $\o^{T,S}$, 
identifying $T$ with $T_+$ and $S$ with $S_+$. To say that $\o$ is 1-distillable
now amounts to requiring that one can choose $S$ and $T$ in such a way that $\o^{T,S}$
approximates $\o_{\rm singlet}$ to arbitrary precision.
 \par
All this applies as yet to the case that $\cA$ and $\cB$ are copies of $B(\bC^2)$.
However, it is not too difficult to generalize everything to the case of a generic
bipartite quantum system. All that needs to be done is to ensure that the 
input state $\o$, defined on the algebra generated by $\cA$ and $\cB$, yields
an output state $\o^{T,S}$ on $B(\bC^2 \otimes \bC^2)$ which can be compared
to $\o_{\rm singlet}$. The formal definition of 1-distillability is then:
\\[6pt]
(4.1) {\bf Definition}
\setcounter{equation}{1}
 Let $\o$ be a state on a general bipartite quantum system
$\cA,\cB \subset B(\cH)$. The state $\o$ is called {\it 1-distillable}
if one can find completely positive maps $T : B(\bC^2) \to \cA$ and $S : B(\bC^2) \to
\cB$ so that the state
$$ \o^{T,S}(x \otimes y) = \o(T(x)S(y))/\o(T({\bf 1})S({\bf 1}))\,, \quad x \otimes y \in B(\bC^2\otimes \bC^2)\,, $$
on $B(\bC^2 \otimes \bC^2)$ approximates $\o_{\rm singlet}$ to arbitrary precision. 
That is to say, given $\epsilon > 0$, there are such $T = T_{\epsilon}$ and
$S = S_{\epsilon}$ so that
\begin{equation} \label{estim}
 |\, \o^{T,S}(X) -\o_{\rm singlet}(X) \,| < \epsilon ||X||\,, \quad X \in B(\bC^2 \otimes \bC^2)\,.
\end{equation}
This criterion for 1-distillability is now completely general and can, in particular, be 
applied in the context of relativistic quantum field theory. This is what we will do
now.
 \par
As in Sec.\ 2, let $(\{\cR(O)\}_{O \subset M},U,\O)$ be a quantum field theory in
vacuum representation. We quote following result, taken from \cite{VerWer}.
\\[6pt]
(4.3) {\bf Theorem} Let $\cA = \cR(O_{\rm A})$ and $\cB = \cR(O_{\rm B})$ be a bipartite quantum
system formed by algebras of local observables localized in spacetime regions
$O_{\rm A}$ and $O_{\rm B}$ which are separated by a non-zero spacelike distance.
Then the vacuum state $\o(\,.\,) = \langle \O,\,.\,\O\rangle$ is 1-distillable on
this bipartite system. Moreover, there is a dense set $\cX \subset \cH$ so that
the vector states $\o_{\chi}(\,.\,) = \langle \chi,\,.\, \chi \rangle$, $||\chi|| =1$,
are 1-distillable on the bipartite system. (In fact, $\cX$ can be chosen so that
this holds for all spacelike separated regions $O_{\rm A}$ and $O_{\rm B}$.)
\\[6pt]
We will add a couple of remarks.
\\[6pt]
(4.4)
The conclusion of the theorem remains valid if one considers the quantum field
theory in a relativistic thermal equilibrium representation instead of a vacuum
representation. Representations of this kind have been introduced by Bros and
Buchholz \cite{BroBu}. The distinction from the vacuum prepresentation is as follows: The 
spectrum condition is dropped, and it is assumed that $\o(\,.\,) = \langle \O,\,.\,\O\rangle$ fulfills  the {\it relativistic KMS condition} at some inverse temperature
$\b > 0$. Following \cite{BroBu}, one says that a state $\o$ on $\cR(\bR^4)$
satisfies the relativistic KMS condition at inverse temperature $\b >0$ (with
respect to the adjoint action of the translation group $U(a)$, $a \in \bR^4$)
if there exists a timelike vector $e$ in $V_+$, the open forward light cone,
so that $e$ hasunit Minkowskian length, and so that for each pair of operators
$A,B \in \cR(\bR^4)$ there is a function $F = F_{AB}$ which is analytic in the
domain $\cT_{\b e} = \{z \in \bC^4 : {\rm Im}\,z \in V_+\cap (\b e - V_+)\}$,
and continuous at the boundary sets determined by ${\rm Im}\, z = 0$, ${\rm Im}\, z = \b e$ 
with the boundary values $F(a) = \langle \O,A U(a) B \O \rangle$,
$F(a +i\b e) = \langle \O,BU(-a)A\O\rangle$ for $a \in \bR^4$. Upon comparison with the
non-relativistic KMS-condition of the previous section, one may get an idea in which way
this is a relativistic generalization of thermal equilibrium states.
\\[6pt]
(4.5)
It is the Reeh-Schlieder theorem which is responsible for the distillability result; we briefly sketch the argument.
In fact, one can show that each non-abelian von Neumann algebra contains an isomorphic
copy of $B(\bC^2)$. In the particular case considered in the situation of Theorem (4.3),
one can use the Reeh-Schlieder theorem to prove that there are algebraic morphisms
$\tau : B(\bC^2) \to \cA$ and $\sigma: B(\bC^2) \to \cB$ so that
$\pi: B(\bC^2 \otimes \bC^2) \to B(\cH)$ given $\pi(x \otimes y) = \tau(x)\sigma(y)$
is a faithful algebraic embedding. Then there is a unit vector $\chi$ in $\cH$
so that $\o_{\chi}(\pi(X)) = \o_{\rm singlet}(X)$ for all $X \in B(\bC^2 \otimes \bC^2)$. According to the Reeh-Schlieder theorem, there is for any
$\epsilon > 0$ some $A= A_{\epsilon} \in \cA$ 
with $||A|| = 1$ so that $||(||A\O||)^{-1}A\O - \chi || < \epsilon$. Thus we choose $T(x) = A^*\tau(x)A$ and $S(y) = \sigma(y)$ to obtain
that the state $\o^{T,S}$ fulfills the required estimate \eqref{estim}.
The Reeh-Schlieder theorem for realativistic thermal equilibrium states has been
proved in \cite{Jae1}.
\\[6pt]
(4.6) The normalization factor $\o(T({\bf 1})S({\bf 1}))$ equals, in the
previous remark, the quantity $\langle \O,A^*\tau({\bf 1})\sigma({\bf 1})A\O\rangle$ which, in turn, is equal to $||A\O||^2$ up to a term of at most order
$\epsilon$. Since we have taken $||A||$ to be equal to 1 (which
made the occurrence of the normalization factor $(||A\O||)^{-1}$
neccessary in the approximation of $\chi$), the quantity
$||A\O||$ here coincides in fact with $||A\O||/||A||$, i.e.\ the effect vs.\
cost ratio which made its appearance in our discussion of the Reeh-Schlieder
theorem. Thus, the factor  $\o(T({\bf 1})S({\bf 1}))$ 
(compared to 1) is a rough
measure for the efficiency of the distillation process,
or put differently, the fraction of members in the  subensemble
corresponding to $\o^{T,S}$ distilled from the
members of the original ensemble $\o$. As we have seen before, this will be
a very small number when $\epsilon$ is small and the spatial distance
between the regions $O_{\rm A}$ and $O_{\rm B}$ is macroscopic for $\o$
the vacuum state.
\\[6pt]
(4.7)
We should like to mention that there are many related works addressing the
issue of long-range correlations in quantum field theory. In fact,
Bell-correlations in quantum field theory have been investigated 
before quantum information theory was established; see the refs.\
\cite{SumWer87a,SumWer,Lan1,Lan2}, and they have contributed to understand
quantum entanglement in a mathematically rigorous form applicable to
general quantum systems. More recent works in this
direction prove that in the bipartite situation $\cA = \cR(O_{\rm A})$,
$\cB = \cR(O_{\rm B})$, for a relativistic quantum field theory, there
is dense set of states violating the Bell-CHSH inequalities \cite{HalCl,Jae3,RRS}.
In this sense, they are quite closely related to the result of the theorem above, which however
gives also information about the distillability of specific states, such as the vacuum or
relativistic thermal equilibrium states, over arbitrarily spacelike subsystems of a
relativistic quantum field theory.
\\[10pt]
I think that, in the light of the theoretical developments
summarized in this contribution, it is fair to say that the interplay between special relativity and quantum physics is holding
a significant position at the frontier of current research. Thus I am quite confident that
special relativity will live well through the next 100 years.

\end{document}